\documentclass[12pt]{iopart}

\usepackage{graphicx,bbm,amssymb,amsopn}

\newcommand{\bra}[1]{{\langle #1 \vert}}

\newcommand{\ket}[1]{{\vert #1 \rangle}}

\newcommand{\ave}[1]{{\langle #1\rangle}}

\newcommand{\ii}{ {\rm i} }

\newcommand{\RR}{\mathbb{R}}
\newcommand{\CC}{\mathbb{C}}

\def\tr{{{\rm tr}\,}}
\def\one{\mathbbm{1}}

\def\one{\mathbbm{1}}

\begin{document}

\title[The two-body random spin ensemble and a new type of quantum phase transition]
{The two-body random spin ensemble\\
and a new type of quantum phase transition}

\author{Iztok Pi\v zorn$^1$, Toma\v z Prosen$^1$, Stefan Mossmann$^2$ and Thomas H. Seligman$^{2,3}$}

\address{
${}^1$ Department of Physics, FMF, University of Ljubljana, Jadranska 19, SI-1000 Ljubljana, Slovenia\\
${}^2$ Instituto de Ciencias F\'\i sicas, Universidad Nacional Aut\'onoma de M\'exico, C.P. 62132 Cuernavaca, Morelos, Mexico\\
${}^3$ Centro Internacional de Ciencias, Apartado postal 6-101, C.P.62132 Cuernavaca, Morelos, Mexico
}

\date{\today}

\begin{abstract}
We study the properties of a two-body random matrix ensemble for distinguishable spins.
We require the ensemble to be invariant under the group of local transformations and
analyze a parametrization in terms of the group parameters and the remaining parameters
associated with the ``entangling" part of the interaction.
We then specialize to a spin chain with nearest neighbour
interactions and numerically find a new type of quantum phase transition related to the
 strength of a random external field i.e. the time reversal breaking one body interaction term.
\end{abstract}


\maketitle

\section{Introduction}

Eugene Wigner introduced random matrix models about fifty years ago into nuclear
physics \cite{wigner}.
The scope of applications has  increased over the years \cite{AT,NT} including
fields such as molecular and atomic physics, mesoscopics and field theory.
More recently random matrix theory has started to be used in quantum information
theory \cite{GS-JQO,GPS-NJP-fidelity,FFS,PS-PRA, GPS-NJP}. For an introduction
to such applications see \cite{PS-AIP}. There the concept of individual qubits
and their interactions becomes important. This implies that we enter the field
of two-body random ensembles (TBRE) \cite{BW-review,FMS-PRE}, i.e. ensembles of
Hamiltonians of $n$-body systems interacting by two-body forces. While such
ensembles have received considerable attention, it was first focussed
on fermions and later also included bosons. Yet in quantum information
theory the qubits are taken to be distinguishable, and indeed the same holds
for spintronics. Interest in both fields has sharply increased recently \cite{NC,spintronix}.

It is thus very pertinent to formulate and investigate TBRE's for distinguishable qubits.
As random matrix ensembles are mainly determined by their symmetry properties this ensemble
will be very different from other TBRE's. In particular, as the particles are distinguishable,
their interaction can vary from particle pair  to particle pair and can indeed be randomly
distributed, thus introducing an entirely new aspect. 
This has the consequence
that the topology according to which spins or qubits are distributed or interact
will be important, Thus chains, trees and crystals of particles with nearest, second
nearest and up to $k$th order interaction can be represented.

As mentioned above, random matrix  ensembles are usually basically defined by
the invariance group of their measure and, if that is not enough, some minimal
information conditions \cite{balian,PS-AIP} or independence condition \cite{Kramers}.
Note that we deal with a symmetry of the ensemble, rather than with a symmetry
of individual systems. The two concepts are to some degree complementary,
and the former has also been called {\em structural invariance}.
We propose an adequate definition for such ensembles in a very general framework in terms of
independent Gaussian distributed variables. We then give an alternate representation
in terms of the invariance group and variables that determine the orbits of the Hamiltonian on the
ensemble under the action of the group.

In order to show the relevance of the new ensemble we address the simplest possible
topology, namely the chain with nearest neighbour interactions. For this system we
focus on the ensemble averaged
structure of the ground state and  demonstrate the existence of an unusual
quantum phase transition \cite{Sachdev},
which is triggered by breaking of
{\em time-reversal invariance} (TRI).

Entanglement, a key resource of quantum many-body systems in terms of quantum information,
is to large extent related to quantum correlations, localization properties and quantum chaos.
Entanglement has also been used as a property, alternative to
long-range order in spatial correlation functions, to describe systems undergoing a quantum
phase transition \cite{Osterloh:02}.
In one-dimensional systems such as quantum spin chains, it was shown \cite{Kitaev} that
the entanglement entropy of the ground state typically saturates or diverges logarithmically
with size when approaching the thermodynamic
limit. Furthermore, it has been shown that logarithmic
divergence implies quantum criticality.

Interesting results emerge when a spatially homogeneous spin model is replaced by
its disordered counterpart, where the spin interactions are taken at random. In this case
there is often no physical justification why random interactions should still obey
specific restricted forms such as Ising or Heisenberg interactions. In this context, we argue,
it is more natural to use
two-spin random ensembles (TSRE) for distinguishable particles, specifically choosing
quantum spins $1/2$, though these ensembles can readily be generalized to arbitrary spin.
By construction these ensembles, as given in section \ref{sec:def},
are invariant with respect to arbitrary local rotations, which we may view as
gauge transformations.
Another physical motivation for the definition of such ensembles is the
coupling among arbitrary and perhaps mutually independent
two level quantum systems which may come from completely different physical contexts
such as e.g. two-level atoms, Josephson junctions and photons.

In section \ref{sec:results} we concentrate on one-dimensional systems or spin chains
and present results of numerical calculations,
mainly based on density matrix
renormalization group (DMRG) \cite{White}, in which we investigate
entanglement and correlation properties of the ground state,
averaged over an ensemble, and the average spectral gap to the first excited state
as well as its fluctuations. If we
include the interaction with an external random magnetic field, and hence TRI is broken,
we find fast decay of correlations, saturation of entanglement entropy, and power law
decay $g\approx N^{-0.4}$
of the spectral gap $g$ with the system size $N$ while its distribution displays
Wigner-type level repulsion.
When the strength of external field goes to zero, and time-reversal
invariance is restored, we find long range
order, logarithmically divergent entanglement entropy, and {\em exponential} decay of
the spectral gap, while the level repulsion disappears.

We argue that this quantum phase transition is non-conventional from the
point of view of established models,
since in what we shall call non-critical case we still find slow power law closing
of the spectral gap.

\section{The embedded ensemble of spin Hamiltonians with random two-body interactions}
\label{sec:def}

In this section we define the two-spin random ensembles of Hamiltonians
for systems with $N$ distinguishable spins or qubits
with at most two body interactions
and describe its basic (invariance) properties. If we do not allow all
spins to interact, we have
to define which ones do. The simplest case will be a chain with nearest neighbour interactions,
but in general we need a graph, whose vertices correspond to spins and whose edges correspond
to two-body interactions. We proceed to formalize this.

Let ${\cal G}=({\cal V},{\cal E})$ be an {\em undirected graph} with a finite set of $N$
{\em vertices} ${\cal V}$ and a set of
$M$ {\em edges} ${\cal E}\subset {\cal V}\times{\cal V}$.
In addition, let $\lambda : {\cal V}\to \RR^+$ and $\mu : {\cal E}\to \RR^+$ be two non-negative
functions defined on the sets of vertices and edges, respectively.
To such a graph we assign a $2^N$ dimensional Hilbert space of $N$ spins or qubits, placed
at its vertices
${\cal H}_{\cal G} = \otimes_{j\in{\cal V}} \CC^2 \equiv \CC^{2^N}$, 
and a set of $N$ Pauli operators
$\sigma^\alpha_j : {\cal H}_{\cal G}\to {\cal H}_{\cal G}, \alpha\in\{1,2,3\},j\in{\cal V}$
satisfying ${\cal SO}(3)$ commutation relations
$[\sigma^\alpha_j,\sigma^\beta_k]=\ii \varepsilon_{\alpha\beta\gamma}\sigma^\gamma_j
\delta_{j,k}$.
We also use the notation $\vec{\sigma}_j = (\sigma^1_j,\sigma^2_j,\sigma^3_j)$.

Let $A^{(j,k)} \in \RR^{3\times 3}$, $(j,k)\in {\cal E}$, be a set of $M$
{\em random} real $3\times 3$ matrices,
and $\vec{b}^{(j)}\in\RR^3$, $j \in {\cal V}$, be a set of $N$ {\em random}
$3$ dimensional real vectors.  The TSRE then consists of the {\em random}
Hamiltonians
\begin{equation}
H = \sum_{(j,k)\in{\cal E}} \mu(j,k)\, \vec{\sigma}_j \cdot A^{(j,k)}
\vec{\sigma}_k + \sum_{j\in{\cal V}}\lambda(j)\, \vec{b}^{(j)} \cdot\vec{\sigma}_j.
\label{eq:H}
\end{equation}
The above defined functions on the edges and vertices of the graph are used to determine
the average strength of the corresponding terms in the Hamiltonian.
The distribution of random two-body interaction matrices (for short also
{\em bond matrices}) $A^{(j,k)}$  and the
random external field vectors $\vec{b}^{(j)}$ shall be uniquely determined
by requiring the following two conditions: {\em maximum local invariance} and {\em maximum
independence} expressed formally as:
\begin{enumerate}
\item An ensemble of Hamiltonians (\ref{eq:H}) should be invariant with
respect to an arbitrary {\em local} ${\cal SO}(3)$ transformation,
namely
\begin{equation}
\vec{\sigma}'_j = O_j \vec{\sigma}_j,
\label{eq:can}
\end{equation}
where $O_j\in {\cal SO}(3)_j$, $ j\in {\cal V}$, meaning that the choice of local coordinate system is arbitrary for each spin/qubit.
Obviously, (\ref{eq:can}) preserves the canonical commutation relations for the Pauli operators.
Then it follows immediately that the {\em joint probability distributions}
of $\{ A^{(j,k)}, \vec{b}^{(j)}\}$ should be invariant with respect to
transformations
\begin{equation}
A^{(j,k)'} = O A^{(j,k)} O', \qquad \vec{b}^{(j)'} = O \vec{b}^{(j)},
\end{equation}
where $O,O'$ are arbitrary independent ${\cal SO}(3)$ rotations for each $j,k$.
\item  The matrix elements of the tensors $A^{(j,k)}_{\alpha,\beta}$ and of the vectors
$\vec{b}^{(j)}_{\alpha}$ should be
{\em independent} random variables.
\end{enumerate}
Following arguments similar to those presented in \cite{Kramers}
it is straightforward to show that, in order to satisfy conditions (i) and (ii) above
for pre-determined but general strengths of bonds $\mu(j,k)$ and external fields $\lambda(j)$,
 $A^{(j,k)}_{\alpha,\beta}$ and $\vec{b}^{(j)}_{\alpha}$  should be
{\em Gaussian independent random variables} of zero mean
 and equal variance, which are uniquely specified in terms of the correlators
\begin{eqnarray*}
\ave{A^{(j,k)}_{\alpha, \beta} A^{(j',k')}_{\alpha',\beta'}} &=&
\delta_{jj'}\delta_{kk'}\delta_{\alpha\alpha'}\delta_{\beta\beta'},\\
\ave{b^{(j)}_\alpha b^{(j')}_{\alpha'}} &=& \delta_{jj'}\delta_{\alpha\alpha'},\\
\ave{A^{(j,k)}_{\alpha,\beta} b^{(j')}_{\alpha'}} &=& 0,
\end{eqnarray*}
where $\ave{\bullet}$ denotes an ensemble average.
We abbreviate the ensemble defined in this way by ${\rm TSRE}({\cal G},\mu,\lambda)$
noting that it depends on the graph and the strength functions $\mu$ and $\lambda$.

Let us now describe some other elementary properties of TSRE.
We have seen that each member $H$ of
 ${\rm TSRE}({\cal G},\mu,\lambda)$ can be parametrized (\ref{eq:H})
by $9 M + 3 N$ independent random parameters.
However, we know for the classical ensembles, that a parametrization in terms of the
structural invariance group, i.e. the invariance group of the ensemble and the remaining
parameters is very useful. Similarly in the present case
for an arbitrary set of local ${\cal SO}(3)$ rotations $O_j$, the transformation
of the parameters
\begin{equation}
A^{(j,k)'} = O_j^T A^{(j,k)} O_k, \qquad \vec{b}^{(j)'} = O^T_j \vec{b}^{(j)},
\label{eq:gauge}
\end{equation}
preserves the spectrum and all entanglement properties of $H$.
In fact, the transformation (\ref{eq:gauge}) can be considered as {\em a gauge
transformation} since,
composed with the local canonical transformation (\ref{eq:can}), it preserves the Hamiltonian
$H$ (\ref{eq:H}) exactly.
Two Hamiltonians, specified by $\{A^{(j,k)},b^{(j)}\}$ and $\{A^{(j,k)'},b^{(j)'}\}$,
can thus be considered equivalent,
and the gauge transformation (\ref{eq:gauge}) defines a natural equivalence relation in
${\rm TSRE}({\cal G},\mu,\lambda)$.
Therefore, it may be of interest to consider a simplest parametrization of the set of
equivalence classes, or in other
words the orbits of a Hamiltonian on the ensemble under the structural invariance transformations.
We thus ask, what is a general {\em canonical form} to which each element $H$
can be brought by gauge transformations (\ref{eq:gauge}) and how can it be parametrized?
Let the integer $K$ denote the number of such
parameters. The equivalent question for the classical ensembles leads to the eigenvalues
as canonical parameters, since there the structural invariance group is much bigger.

Let us first consider the simplest connected graph, namely an open one dimensional
(1D) chain of $N$ vertices
$\{ 1,\ldots N\}$ and $M=N-1$ bonds.
There it turns out that the matrices $A^{(j,j+1)}$ can be simultaneously
{\em symmetrized}, namely all
$A^{(j,j+1)'}=[A^{(j,j+1)'}]^T$, by choosing the following gauge transformation
\begin{equation}
O_{j+1} = R_j O_j,\quad \textrm{where}\quad
R_j:=V^{(j,j+1)} [U^{(j,j+1)}]^T
\end{equation}
and where
 \begin{equation}
 A^{(j,j+1)}=: U^{(j,j+1)} D^{(j,j+1)} [V^{(j,j+1)}]^T,
 \label{eq:SVD}
 \end{equation}
 is a standard canonical singular value decomposition (SVD) of the original
bond matrix $A^{(j,j+1)}$,
 with $U^{(j,j+1)},V^{(j,j+1)}\in {\cal SO}(3)$ and $D^{(j,j+1)}$ diagonal
matrices of singular values.
 Since the initial transformation $O_1$ is still free, we can choose it such $O_1:= U^{(1,2)}$
 that the first bond matrix is even {\em diagonalized}, $A^{(1,2)'} = D^{(1,2)}$.
 This symmetrization is unique provided that singular values of all SVD's (\ref{eq:SVD}) are
 non-degenerate which is the case for a generic member $H$. Thus the number
of parameters specifying the
 bond matrices is $3+6(M-1) = 6M-3$ and in addition to $3N$ external
field parameters this gives
 $K=6M-3+3N$ independent parameters. We recover the original set of parameters
if we add the $3N$ parameters
of the group of local rotations.

Second, we consider the case of a ring graph with $N$ vertices and $M=N$ bonds,
which is obtained from the previous case by
specifying the periodic boundary condition $N+1\equiv N$. We see that a general $H$
as given in (\ref{eq:H}) can
now be symmetrized only if the additional topological condition $R:=R_N \cdots R_2 R_1 = \one$
is satisfied.
Now all but one bond matrix can be symmetrized, for example the last one may in
addition be multiplied by a topological
rotation $A^{(N,1)'} = A^{(N,1)'}_{\rm symmetric} R$ leading to three additional parameters.
Therefore in the case of a ring graph we have
$K=6M+3N$ independent parameters; again adding the $3N=3M$ parameters of the local
rotations we obtain the full set of parameters.

We can now consider the case of a general (connected) graph. From the previous examples
it is evident that the only crucial
additional parameter is the number $L$ of {\em primitive} cycles, i.e. such cycles which
cannot be decomposed into other primitive cycles.
It is clear that each primitive cycle adds $3$ additional topological parameters
(or one ${\cal SO}(3)$
topological $R$ matrix) to the $6M-3+3N$ parameters which we
would have for the case of a {\em tree} graph. Therefore we have
\begin{equation}
K = 6M + 3N + 3L - 3.
\end{equation}
Counting the cycle-contributions and taking the primitivity criterion into account again
we obtain the the total number of parameters by adding those of the local rotations.

The above considerations hold for the case of general bond and vertex
strength functions $\mu$ and $\lambda$.
If, however, these functions are degenerate, or even constant, i.e. the average interaction
strength and field strength do not
depend on edges/vertices of the graph, then
the structural invariance group of the TSRE may be even larger.
In the latter case this group is obtained as a semi-direct product of
a {\em discrete symmetry group} of the graph ${\cal G}$ and the
gauge group of local rotations, the latter being the normal subgroup.
We hope that the considerable invariance properties of the TSRE will prove useful in
a future analytical treatment of its properties.

\section{Properties of ground states of the TSRE on a 1D chain}

\label{sec:results}

In this section we shall only consider the simplest case of a TSRE on a 1D chain of
$N$ vertices. In addition, we consider the
most symmetric case of constant strength functions, say $\mu(j,j+1) \equiv 1$ and $\lambda(j)\equiv \lambda={\rm const}$. Such an ensemble of random spin chain
Hamiltonians
\begin{equation}
     H = \sum_{j=1}^{N-1} \vec{\sigma}_j \cdot A^{(j,j+1)} \vec{\sigma}_{j+1} + \lambda
\sum_{j=1}^N \vec{b}^{(j)} \cdot\vec{\sigma}_j
\label{eq:H1}
\end{equation}
shall be designated as ${\rm TSRE}(N,\lambda)$ where we explicitly assume open
boundary conditions; we shall, however,
also consider a ring graph with periodic boundary conditions in which case we shall
stress this separately.
In particular we shall be interested in the zero temperature (ground state) properties of
${\rm TSRE}(N,\lambda)$. We note that due to the large co-dimension of bond strength
space the standard perturbative renormalization group of
decimating the strongest bonds \cite{Fisher} would not work and we have at this point to
rely on a brute numerical investigation.

Still, it turns out that most zero temperature properties of ${\rm TSRE}(N,\lambda)$ can be
efficiently simulated using White's density matrix renormalization
group (DMRG) finite-size algorithm \cite{White}
by which spin chains of sizes up to $N=80$ could at present be achieved.
One should not forget that
all numerical estimates of {\em ensemble average} or expectation value of some
physical quantity $A$, which  will be designated as $\ave{A}$, require averaging over many,
say ${\cal N}_{\rm r}$, realizations from ${\rm TSRE}(N,\lambda)$, such that the statistical error
estimated as $\sigma_A \sim\sqrt{(\ave{A^2}-\ave{A}^2)/{\cal N}_{\rm r}}$ is sufficiently small.
Due to the lack of translational invariance the implementation
of the DMRG is non-trivial and was only done for a chain with open boundaries.

We note that for $\lambda=0$, any $H$ as defined in (\ref{eq:H1}),
or even more generally in (\ref{eq:H}), commutes with the following anti-unitary
{\em time-reversal} operation
\begin{equation}
\hat{T} : \vec{\sigma}_j \to -\vec{\sigma}_j,\qquad H\vert_{\lambda=0}\, \hat{T} = \hat{T}\, H\vert_{\lambda=0},
\end{equation}
so, for {\em odd} $N$, all eigenvalues of $H$ have to be doubly degenerate
(Kramer's degeneracy \cite{Kramers}). However, as this represents more a
technical than conceptual
problem for the ground state (or ground plane) properties of
${\rm TSRE}(N,\lambda)$ we shall in the
following restrict ourselves to the case  of {\em even} $N$.

\subsection{Distribution of the spectral gap}
Let $\ket{0},\ket{1}$, represent the ground state, and the first excited state,
of $H$,
with eigenenergies $E_0$, and $E_1$, respectively. It is well known that the crucial quantity
which determines the rate of relaxation of zero-temperature quantum dynamics is the spectral
gap $g = E_1 - E_0$.

\begin{figure}[h!]
\center{\includegraphics[width=0.8\textwidth]{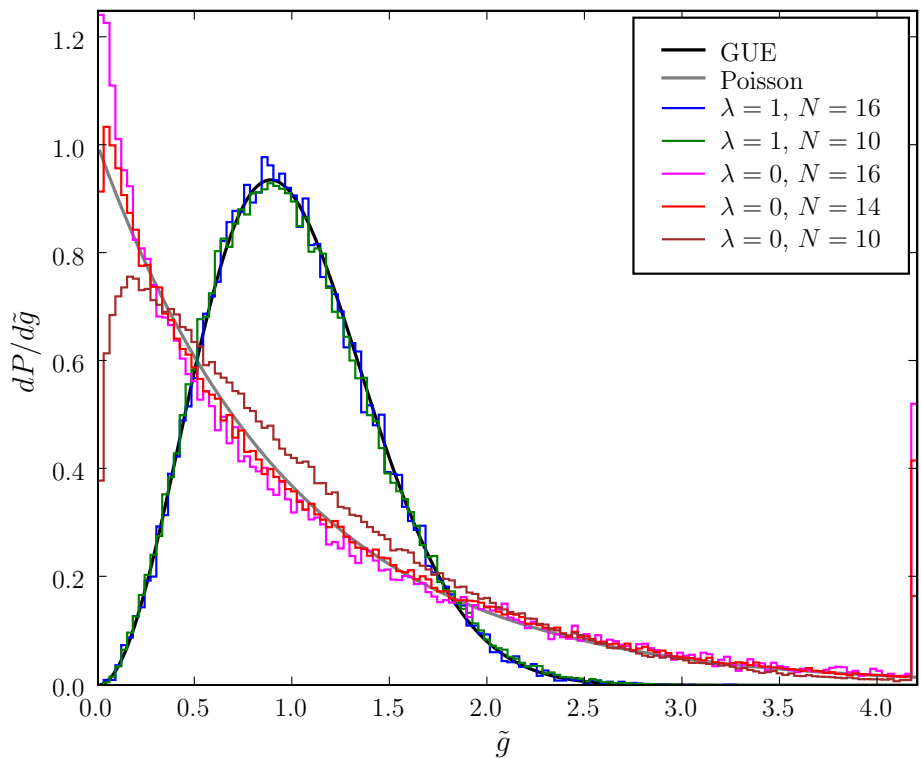}}
\caption{
Distribution of normalized gap ${\tilde g}=g/\ave{g}$ for a few choices of even $N$.
In the non-$\hat{T}$-invariant case an agreement with GUE level spacing distribution
is obtained whereas
in the $\hat{T}$-invariant case, $\lambda=0$, the level repulsion gradually vanishes
in the thermodynamic limit.
}
\label{fig:1}
\end{figure}

To clarify the difference with respect to
TRI, we plot the normalized gap distribution
$dP/d{\tilde g}$ where ${\tilde g} = g/\ave{g}$ for both cases,
together with the theoretical level spacing distributions for
the Gaussian unitary ensemble (GUE) of random matrices \cite{mehta}
and  for an uncorrelated Poissonian spectrum (Fig.~\ref{fig:1}).
For the non-TRI case we choose $\lambda=1$
and observe, to our accuracy, good agreement with
the GUE case for two chain sizes $N=10$ and $N=16$.
Our results suggest that the GUE-like gap distribution, exhibiting level repulsion,
also holds in the thermodynamic limit $N\to\infty$;
numerical results for odd $N$ give the same results.
In the case of $\lambda=0$ the level repulsion between $\ket{0}$ and $\ket{1}$
gradually vanishes as we approach the thermodynamic limit, although
no conclusive statement can be made about the limiting distribution.
For odd $N$, the ground state is degenerate for $\lambda=0$
and the present analysis does not apply.

\subsection{Size scaling of the spectral gap}

\begin{figure}[h!]
\center{\includegraphics[width=0.8\linewidth]{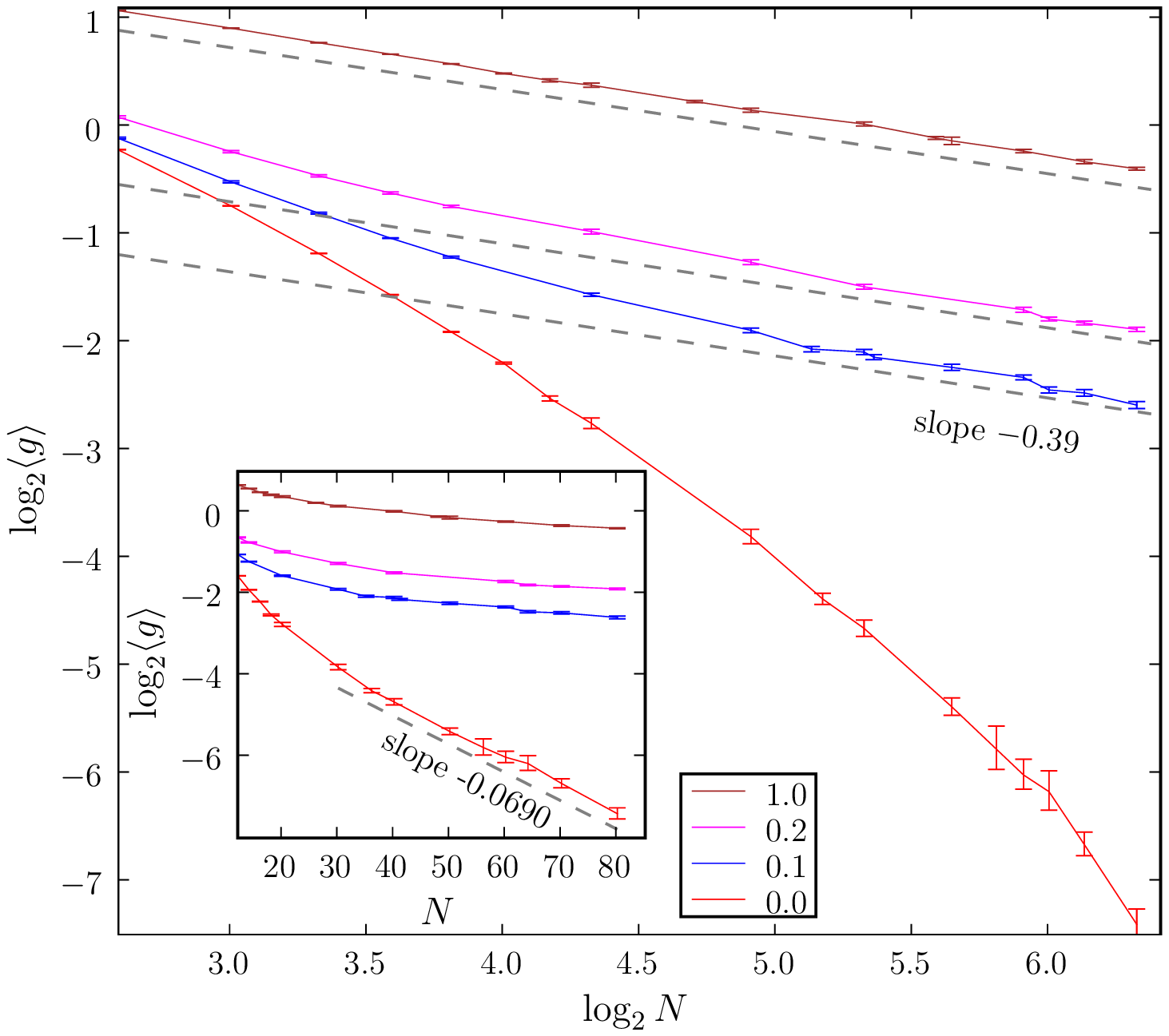}}
\caption{
Spectral gap scaling with the system size for different values of parameter $\lambda$ (see legend).
Note an asymptotic scaling $\propto N^{-0.39}$, unless $\lambda=0$ where faster than power
law decay of a gap is observed, perhaps asymptotically exponential (see inset for a semi-log scale).
}
\label{fig:2}
\end{figure}

Being interested in the thermodynamic limit, it is an
important issue to understand how $g(N)$ scales with $N$. The theory of quantum criticality
\cite{Sachdev} states that $g$ remains finite in the thermodynamic limit for {\em non-critical systems},
and rapidly
converges to zero, as $N\to\infty$ for {\em critical} systems.

In Fig.\ref{fig:2} we plot the ensemble averaged spectral gap $\ave{g}$
versus $N$ for different values
of the field strength $\lambda$. We find a clear indication that in the non-TRI case
the spectral gap exhibits universal asymptotic power law scaling
\begin{equation}
\ave{g} \sim N^{-\eta},\quad \textrm{with} \quad \eta \approx 0.39 \pm 0.01,
\label{eq:powerlaw}
\end{equation}
whereas in the TRI case, $\lambda=0$, the asymptotic decay of the gap is {\em faster
than a power law}, perhaps exponential $\ave{g} \sim \exp(-\xi N)$,
with $\xi \approx 0.07\pm 0.02$.
According to the standard theory \cite{Sachdev} both cases,
$\lambda\neq 0$ and $\lambda = 0$, should
be classified as quantum critical, however as we shall see later, the case of slow
power-law decaying average gap (\ref{eq:powerlaw}) has many-features of non-critical systems, such as
finite correlation length and finite (saturated) entanglement entropy.
Therefore we shall, at least for the
purposes of the present paper, name the case $\lambda \neq0$ as {\em random non-critical}
(RNC) and
the case $\lambda = 0$ as {\em random critical} (RC).

We note that the results for odd and even $N$ are in agreement
in the RNC case.
Also results for the case with periodic boundary conditions up to $N=20$
show no significant difference from the results in Fig.~\ref{fig:2}
for any $\lambda$.

\subsection{Size scaling of the ground state entanglement entropy}

\begin{figure}[h!]
\center{\includegraphics[width=0.8\linewidth]{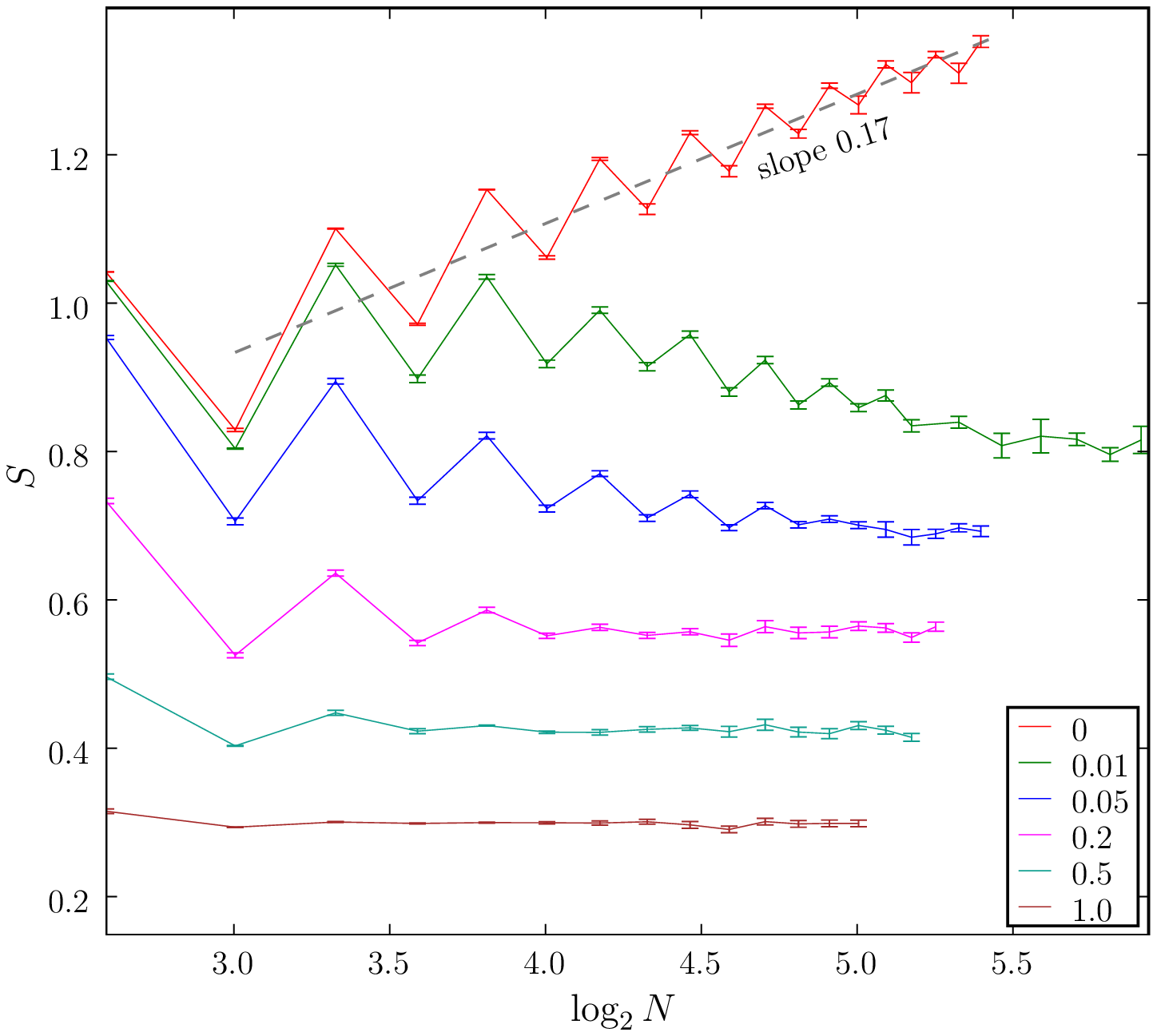}}
\caption{
Entanglement entropy versus chain length $N$ for different values of parameter $\lambda$
(see legend).
Logarithmic divergence with an estimated asymptotic slope $S \sim 0.17 \log_2 N$
for the critical case $\lambda=0$ is indicated with a dashed line.}
\label{fig:3}
\end{figure}

The second characteristics of quantum phase transitions we choose to investigate in the
${\rm TSRE}(N,\lambda)$, is the entanglement entropy of a symmetric bi-partition of the chain
\begin{equation}
S(N,\lambda) = \ave{\tr_{\{N/2+1,\ldots N\}}\left[ (\tr_{\{1,\ldots,N/2\}} \ket{0}\bra{0})
 \log_2 (\tr_{\{1,\ldots,N/2\}} \ket{0}\bra{0})\right]},
\end{equation}
which measures the entanglement in the ground state between two equal halves of the chain.
It has been suggested in non-random systems \cite{Kitaev} that for critical cases
$S \propto \log_2 N$ whereas in non-critical cases $S$ saturates in the thermodynamic limit.

Indeed, as shown in Fig.\ref{fig:3}, we find for the ${\rm TSRE}(N,\lambda)$ that $S$ saturates
to a constant finite $S_\infty(\lambda) = \lim_{N\to\infty}S(N,\lambda)$ for the RNC
case $\lambda\neq 0$,
while in the RC case it grows logarithmically
\begin{equation}
S(N,\lambda=0) \approx c \log_2 N + c', \quad {\rm with}\quad c\approx 0.17\pm 0.02
\end{equation}
We also note an interesting even-odd-$N/2$ effect which slowly diminishes
as we approach the thermodynamic limit.
As pointed out in Ref. \cite{laflorenciePRL96} such an effect is induced by
open boundary conditions.
For periodic boundary conditions the entanglement entropy
for RNC case is twice as large as in the ${\rm TSRE}(N,\lambda)$ with open boundaries.
This further
confirms the conjecture that only short-range correlations around the boundary
between the two halves contribute to the entanglement.

We note that our result is essentially different from results for other models, which can be
obtained by perturbative real space renormalization
group \cite{Fisher}, for example for the disordered critical Heisenberg chain \cite{rafel},
where $c = (\ln 2)/3$ and
is in general model dependent \cite{santachiara}.

The fact that the entanglement is reduced in the RNC case with local disorder
can be explained by chaotic behavior \cite{Lea:04} signalized by the
level repulsion in the gap distribution. A similar effect
can be observed in localization properties where hopping of excitations
induced by inter-particle interactions is diminished by introducing
local disorder \cite{Dykman:04,Olivier:07}.
This effect of increased localization is useful for successful quantum computing
and has important consequences for transport properties such as conductivity \cite{Altshuler}.

%{\bf IP: } entanglement, localization and simulability? \cite{Montangero:04}

\subsection{Correlation functions}

\begin{figure}[h!]
\center{\includegraphics[width=0.8\linewidth]{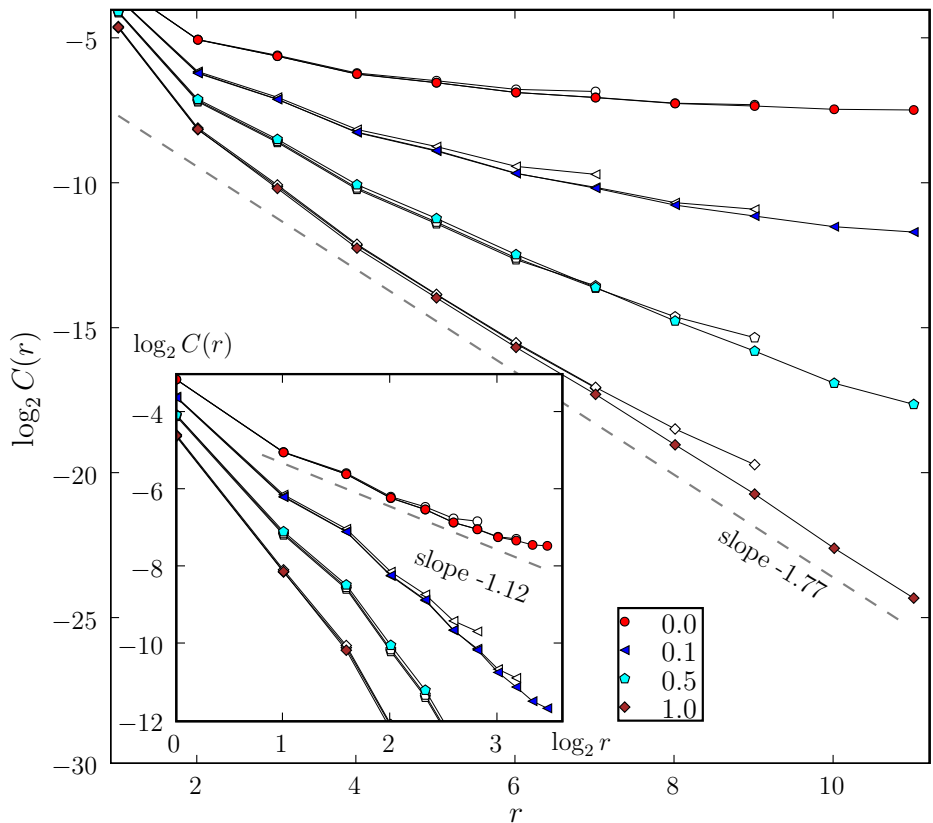}}
\caption{
Ensemble fluctuations of the ground state spin-spin correlation function $C(r)$ versus
distance $r$, for different values of the parameter $\lambda$ (indicated in the legend).
Open symbols indicate results for chain length $L=16$ and $L=20$ whereas
closed symbols stand for $L=24$.}
\label{fig:4}
\end{figure}

The most direct probe of criticality is perhaps to investigate of long-range order and
(space) correlation functions. In order to do this we compute the ensemble averaged
fluctuation of the spin-spin correlation function between two vertices
\begin{equation}
C(j,k) =
\ave{|\bra{0}\sigma_j^\alpha \sigma_k^\beta\ket{0} -
\bra{0}\sigma_j^\alpha\ket{0}\bra{0}\sigma_k^\beta\ket{0}|^2}.
\label{eq:C}
\end{equation}
Note that we have to consider average fluctuations
of the spin-spin correlation function instead of the correlation function itself, since
the latter have to vanish due to the local gauge invariance properties of the TSRE.
Because of the
local invariance of the TSRE it is enough to consider a single type of correlation function,
as the RHS of (\ref{eq:C}) {\em does not} depend on indices $\alpha,\beta$ if $j\neq k$;
in fact
in numerical computations we average over $\alpha,\beta$ in order to improve statistics.
We consider the Hamiltonian (\ref{eq:H1}) with
periodic boundary conditions. This allows us to average the
fluctuation of the correlation over the chain and hence
     $C(r) = \frac{1}{N} \sum_i C(i,i+r)$.
We expect that the results would be qualitatively the same for the
model with open boundaries and sufficiently large $N$, but we obtain better statistics in this way.

%%%%%%%%%%%%%5
Fig.~\ref{fig:4} shows the averaged correlation function fluctuation $C(r)$
for a few choices of the control parameter $\lambda$ and of the chain lengths $N=16,20,24$.
The results for chains of different lengths coincide for small
distances $r$ whereas for larger $r$ finite-size effects are noticeable.
For sufficiently large chains it can be conjectured that
the fluctuations (or effective correlations) asymptotically decay in the RNC case
as $C(r) \asymp C_0 2^{-r/\xi}$ with a finite correlation length $\xi$
whereas the decay for the RC case is slower than exponential, perhaps a power-law, which
indicates long-range order $\xi=\infty$.

\subsection{Correlation length and the entanglement entropy saturation value}

In Fig.~\ref{fig:4} we observe that the degree of localization depends
on the control parameter $\lambda$ and the correlation function $C(r)$ decays
on larger scales as we approach the critical point $\lambda=0$
which results in a larger correlation length $\xi$.
Eventually, the correlation length becomes infinite at the critical point $\lambda=0$.
\begin{figure}[h!]
\center{\includegraphics[width=0.85\linewidth]{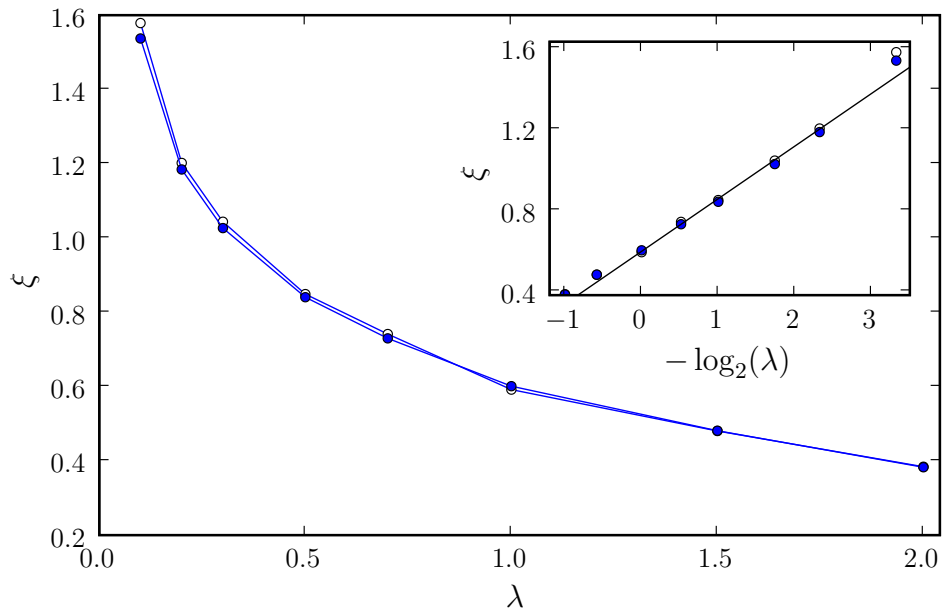}}
\caption{The correlation length $\xi$ as a function of the
control parameter $\lambda$.
Open and full symbols designate correlation length obtained from the best of $C(r)$ with $c \exp(-r/\xi)$ on sets
$r\in \{4,\ldots,8\}$ and $r\in \{5,\ldots,7\}$, respectively, all for $N=24$.
The inset demonstrates $\propto -\log_2\lambda$ scaling. 
Please
note that last point at $\lambda=0.1$ is rather inaccurate due to insufficiently
large system size $N$, so it is not used in a linear fit (full line) in the inset, for which
other points with $\lambda \le 1$ have been used.
}
\label{fig:5}
\end{figure}
In Fig.~\ref{fig:5} we show the dependence of the correlation length $\xi$ on
the control parameter $\lambda$ as obtained by exponential fit
of $C(r)$ for a finite size $N=24$. 
Unlike conventional phase transition as in e.g. \cite{Osterloh:02},
the correlation length seems to diverge logarithmically as
$\xi(\lambda) \sim -\xi_0 \log_2 \lambda + {\rm const}$ with $\xi_0 =  0.26$ (indicated in the
inset of fig.\ref{fig:5}),
even though algebraic scaling cannot be entirely excluded with the numerical
data that are available at present.

\begin{figure}[h!]
\center{\includegraphics[width=0.85\linewidth]{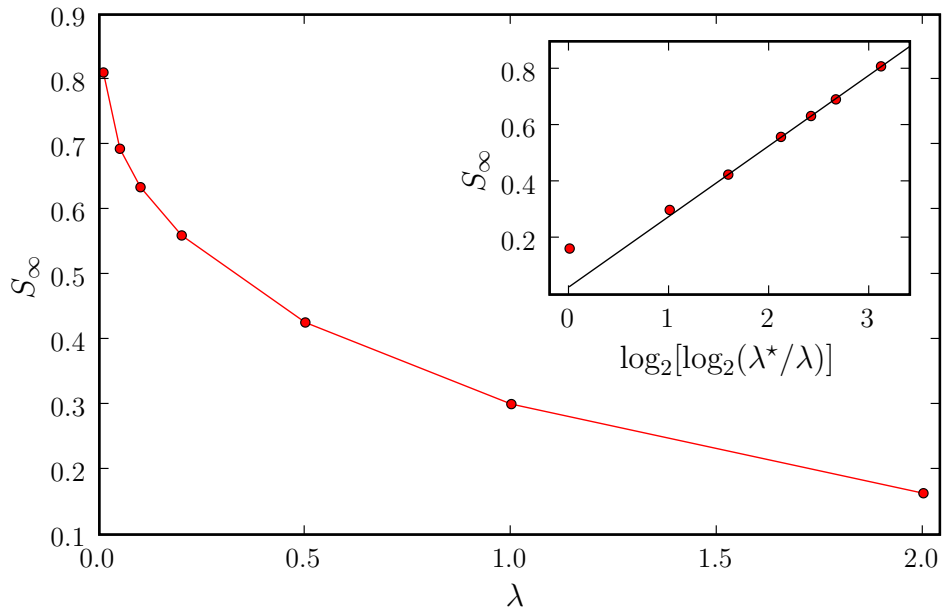}}
\caption{Saturated entanglement entropy $S_\infty$
as a function of the control parameter $\lambda$. The inset demonstrates
$\propto \log_2 \log_2(\lambda^\star/\lambda)$ scaling, where $\lambda^*=4$, which is indicated by a straight line
obtained from a best fit to points with $\lambda < 1$. 
}
\label{fig:6}
\end{figure}

Large correlation length has strong effect on entanglement.
Long range correlations demand longer chains for the entanglement entropy
to saturate whereas the saturation value itself also grows when
the critical point is approached.
In Fig.~\ref{fig:6} we plot
the entanglement entropy saturation value for various values of parameter $\lambda$
and observe similar behavior as for the correlation length.
However, unlike the correlation length the quantity which diverges logarithmically
when $\lambda \to 0$ is not the entanglement entropy $S$ but its exponential.
In fact, numerical data for small $\lambda$ show good agreement with $2^{k S} \propto -\log_2 \lambda + {\rm const}$, where
$k = 4.0$ (indicated in the inset of fig.\ref{fig:6}).
Note that $2^S$ is to a good approximation proportional to the effective rank,
or the Schmidt number of the
ground state $\chi^{\epsilon}$ \cite{Kitaev,MPS}, which denotes
the number of eigenvalues of the reduced density matrix
needed to describe the state of the system up to an error $\epsilon$.
%according to a bipartition of the chain into two halves.
In fact, the effective rank $\chi^\epsilon$, rather than the entanglement entropy,
is the decisive indicator of simulability by the DMRG method \cite{Schuch:07}
and, we believe, also a relevant quantity in the description of
a quantum phase transition.

%{\bf IP:} localization vs entanglement (Li, Wang, and Hu: quant-ph/0308116).

%{\bf IP:} logarithmic light-cone: \cite{Osborne:07}. But there is no time dependence in our model!

\section{Conclusions}
In the present paper we have defined a two-body random matrix ensemble of
independent spin Hamiltonians which are invariant under local ${\cal SO}(3)$
transformations and described them in a framework of undirected graphs.
As the simplest example, we have studied a chain with nearest-neighbour
interactions in a random external field and observed a non-conventional phase transition
when the external field is switched off. The system is always critical in conventional terminology
as it has a vanishing gap in the thermodynamic limit in all cases studied.
Yet we have shown that, in the presence of a random external field breaking
time-reversal invariance, the locally disordered
system has many properties of non-critical systems such
as finite correlation length and finite bipartite entanglement entropy in the thermodynamic limit,
whereas the gap decay obeys a universal power law dependence.
The transition towards the critical point with vanishing of the external field
exhibits logarithmic divergence
for the correlation length and the effective rank of the ground state.
We have no explanation for the logarithmic behavior in the
quantum phase transition.

The model proposed is much richer than the example discussed. Thus we expect,
that higher connectivity of the graph will yield very different results, but even
an exploration of high temperature behaviour for the chain seems very worthwhile.
In view of the large structural invariance group of the ensemble in the case of
site independent average coupling and external fields we hope,
that some analytic results can be obtained for this ensemble.

%{\bf IP:}
%Something about conductivity and localization? \cite{Altshuler}?
%But I have no idea how this localization in \cite{Altshuler}
%could be related to our model?
%After all, it is related to finite temperatures.

\section*{Acknowledgments}
We acknowledge support by Slovenian Research Agency, program P1-0044, and grant J1-7437,
by CONACyT under grant 57334 and by UNAM-PAPIIT under grant IN112507.
IP and TP thank THS and CIC Cuernavaca for hospitality.


\section*{References}

\begin{thebibliography}{99}

\bibitem{wigner} E. P. Wigner, Ann. Math. {\bf 53}, 36 (1951).

\bibitem{AT} T. A. Brody, J. Flores, J. B. French, P. A. Mello, A. Pandey and S. S. M. Wong, Rev. Mod. Phys. {\bf 53}, 385 (1981).

\bibitem{NT}  T. Guhr, A. M\" uller-Groeling and H. A. Weidenm\" uller, Phys. Rept. {\bf  299}, 189 (1998).

\bibitem{GS-JQO} T. Gorin and T. H. Seligman, J. Quant. Opt. B, 4, S386 (2002)

\bibitem{GPS-NJP-fidelity} T. Gorin, T. Prosen and T. H. Seligman, New J. Phys. {\bf 6}, 20 (2004).

\bibitem{FFS} K. M. Frahm, R. Fleckinger and D. L. Shepelyansky, Eur. Phys. J. D {\bf 29}, 139 (2004).

\bibitem{PS-PRA} C. Pineda and T. H. Seligman, Phys. Rev. A {\bf 75}, 012106 (2007).

\bibitem{GPS-NJP} T. Gorin, C. Pineda and T. H. Seligman, New J. Phys, {\bf 9}, 206 (2007).

\bibitem{PS-AIP} C. Pineda and T. H. Seligman, to be published in ELAF 2007 Proceedings, AIP

\bibitem{BW-review} L. Benet and H. A. Weidenm\" uller, J. Phys. A: Math. Gen. {\bf 36}, 3569 (2003).

\bibitem{FMS-PRE} J. Flores, M. Horoi, M. M\" uller and T. H. Seligman, Phys. Rev. E {\bf 63}, 026204 (2000).

\bibitem{NC}  M. A. Nielsen and I. L. Chuang, \emph{Quantum Computation and Quantum Information}
(Cambridge University Press, 2000)

\bibitem{spintronix} I. Zutic, Rev. Mod. Phys. {\bf 76}, 323 (2004).

\bibitem{balian} R. Balian, Nuovo Cimento {\bf B57}, 183 (1958).

\bibitem{Kramers} F. Haake, \emph{Quantum Signatures of Chaos} (Springer-Verlag, Berlin, 2001).

\bibitem{Sachdev} S. Sachdev, \emph{Quantum Phase Transitions} (Cambridge University Press, Cambridge, 1999).

\bibitem{Osterloh:02} A. Osterloh, L. Amico, G. Falci, and R. Fazio, Nature \textbf{416}, 608 (2002).

\bibitem{Kitaev} G.~Vidal, J.~I.~Latorre, E.~Rico, and A.~Kitaev, Phys. Rev. Lett. \textbf{90}, 227902 (2003).

\bibitem{White} S. R. White, Phys. Rev. Lett. \textbf{69}, 2863 (1992); S. R. White, Phys. Rev. B \textbf{48}, 10345 (1993).

\bibitem{Fisher} C. Dasgupta and S. Ma, Phys. Rev. B \textbf{22}, 1305 (1980);
              D. S. Fisher, Phys. Rev. B \textbf{50}, 3799 (1994).

\bibitem{mehta} M. L. Mehta, \emph{Random matrices} (Academic University Press, 1991).

\bibitem{laflorenciePRL96} N. Laflorencie, E. S. S{\o}rensen, M.-S. Chang, and I. Affleck,
              Phys. Rev. Lett. \textbf{96}, 100603 (2007).

\bibitem{rafel}  G. Refael and J. E. Moore, Phys. Rev. Lett. {\bf 93}, 260602 (2004).

\bibitem{santachiara} R. Santachiara, J. Stat. Mech. 2006, L06002 (2006).

\bibitem{Lea:04} L.~F.~Santos, G.~Rigolin, and C.~O.~Escobar, Phys.~Rev.~A {\bf 69}, 042304 (2004); L.~F.~Santos, J.~Phys.~A {\bf 37}, 4723 (2004).

\bibitem{Dykman:04} M.~I.~Dykman, F.~M.~Izrailev, L.~F.~Santos, and M.~Shapiro, {\tt e-print} cond-mat/0401201.

\bibitem{Olivier:07} O.~Giraud, J. Martin, and B. Georgeot, Phys.~Rev.~A {\bf 76}, 042333 (2007).

\bibitem{Altshuler} D.~M.~Basko, I.~L.~Aleiner, and B.~L.~Altshuler, {\tt e-print} cond-mat/0602510; Annals of Physics {\bf 321}, 1126 (2006).

\bibitem{MPS} G.~Vidal, Phys.~Rev.~Lett. {\bf 91}, 147902 (2003); {\em ibid.} {\bf 93}, 040502 (2004);
 S.~R.~White and A.~E.~Feiguin, Phys.~Rev.~Lett. {\bf 93}, 076401 (2004);
A.~J.~Daley, C.~Kollath, U.~Schollw\"{o}ck, and G.~Vidal, J.~Stat.~Mech. {\bf 4}, P04005 (2005).

\bibitem{Schuch:07} N.~Schuch, M.~W.~Wolf, F.~Verstraete, and J.~I.~Cirac, {\tt e-print} arXiv:0705.0292.


%\bibitem{Montangero:04} S.~Montangero, Phys.~Rev.~A {\bf 70}, 032311 (2004).

%\bibitem{Viola:07} L.~Viola and W.~G.~Brown, {\tt e-print} quant-ph/0702014.

%\bibitem{Osborne:07} C.~H.~Burrell and T.~J.~Osborne, {\tt e-print} quant-ph/0703209.

\end{thebibliography}
\end{document}